# STATISTICAL MECHANICS OF NEOCORTICAL INTERACTIONS:
# REACTION TIME CORRELATES OF THE $g$ FACTOR

Lester Ingber

Lester Ingber Research

PO Box 06440

Wacker Dr PO Sears Tower

Chicago, IL 60606

and

DRW Investments LLC

311 S Wacker Dr Ste 900

Chicago, IL 60606

ingber@ingber.com, ingber@alumni.caltech.edu

http://www.ingber.com/

Psycholoquy Commentary on

The $g$ Factor: The Science of Mental Ability

by Arthur Jensen

ABSTRACT: A statistical mechanics of neuronal interactions (SMNI) is explored as providing some substance to a physiological basis of the $g$ factor. Some specific elements of SMNI, previously used to develop a theory of short-term memory (STM) and a model of electroencephalography (EEG) are key to providing this basis. Specifically, Hick's Law, an observed linear relationship between reaction time (RT) and the information storage of STM, in turn correlated to a RT-$g$ relationship, is derived.

# 1. Introduction

## 1.1. Context of Review

My specific interest in taking on this review of "The *g* Factor" by Arthur Jensen (AJ) is to see if some anatomical and/or physiological processes at the columnar level of neuronal interactions can account for the *g* factor.

From circa 1978 through the present, a series of papers on the statistical mechanics of neocortical interactions (SMNI) has been developed to model columns and regions of neocortex, spanning mm to cm of tissue. Most of these papers have dealt explicitly with calculating properties of short-term memory (STM) and scalp EEG in order to test the basic formulation of this approach. SMNI derives aggregate behavior of experimentally observed columns of neurons from statistical electrical-chemical properties of synaptic interactions. While not useful to yield insights at the single neuron level, SMNI has demonstrated its capability in describing large-scale properties of short-term memory and electroencephalographic (EEG) systematics (Ingber, 1982; Ingber, 1983; Ingber, 1984; Ingber, 1991; Ingber, 1994; Ingber, 1995a; Ingber & Nunez, 1995; Ingber, 1996a; Ingber, 1997).

## 1.2. Errors In Simple Statistical Approaches

I must assume that AJ faced very difficult problems in choosing just how much technical details to give in his broad text, e.g., discussing the extent of expert statistical analyses that have been brought to bear upon the *g* factor. However, I do see reason to criticize some general features of the simple statistical algorithms presented, especially those that overlap with my own mathematical and physics expertise.

The simple approach to factor analysis initiated on page 23,

$$X = t + e, \qquad (1)$$

where *e* is the residual "noise" of fitting the variable *X* to the independent variable *t*, has some serious flaws not addressed by additional material presented thereafter. For example, in this context, I find the arguments in the long footnote 16 on pages 101-103 unconvincing, but I agree with its conclusion:

> But the question is mainly of scientific interest, and a really satisfactory answer ... will become possible only as part and parcel of a comprehensive theory of the nature of *g*. ... The distribution of obtained measurements should conform to the characteristics of the distribution dictated by theoretical considerations.

I think it clear that any such "theoretical considerations" must themselves be well tested against experimental evidence at each spatial-temporal scale purported to be modeled.

It must be understood that there is a quite explicit model being assumed here of the real world — that of a simple normal Gaussian process. The real issue in many physical/biological systems is that most often the real multivariable world is much more aptly described by something like

$$X = t_X(X, Y) + s_X(X, Y) \, e_X \tag{2.a}$$

$$Y = t_Y(X, Y) + s_Y(X, Y) \, e_Y. \tag{2.b}$$

When the *t*'s and *s*'s are constants, then simple statistics can determine their values and cross-correlations between the *s*'s.

Simple statistical methods can even do OK if the *t*'s are relatively simple quasi-linear parameterized functions. Such simple methods fail quite miserably if the *t*'s are highly nonlinear functions, especially if care is not taken to employ sophisticated optimization algorithms. The most terrible flaws often occur because, for the sake of making life easier for the analyst, any model faithful to the real system is butchered and sacrificed, and the *s*'s are taken to be constants. In general, there can be a lot of "signal" in the (generally nonlinear) functionality of the "noise" terms. In general, no degree of fancy quasi-linear statistical analysis can substitute for a proper theory/model of the real system.

In general, the proper treatment of the problem is quite difficult, which is of course no excuse for poor treatment. The solution in many disciplines is to go a level or two deeper in some Reductionist sense, to develop plausible models at the top scale being analyzed. Indeed, this was the call I saw and responded to in the advertisement for reviewers of the work by AJ:

> Commentary is invited from psychometricians, statisticians, geneticists, neuropsychologists, psychophysiologists, cognitive modellers, evolutionary psychologists and other specialties concerned with cognitive abilities, their measurement, and their cognitive and neurobiological basis.

In this context, the successes of SMNI and its agreement with general STM observations are due to processing stochastic nonlinearities of the forms described above. Attempts to avoid dealing with these nonlinearities, derived from lower-level synaptic and neuronal activity, have not been as successful as SMNI in detailing STM (Ingber, 1995b).

## 2.  SMNI Description of Short-Term Memory (STM)

Since the early 1980's, a series of papers on the statistical mechanics of neocortical interactions (SMNI) has been developed to model columns and regions of neocortex, spanning mm to cm of tissue.

Most of these papers have dealt explicitly with calculating properties of short-term memory (STM) and scalp EEG in order to test the basic formulation of this approach (Ingber, 1981; Ingber, 1982; Ingber, 1983; Ingber, 1984; Ingber, 1985a; Ingber, 1985b; Ingber, 1986; Ingber & Nunez, 1990; Ingber, 1991; Ingber, 1992; Ingber, 1994; Ingber & Nunez, 1995; Ingber, 1995a; Ingber, 1995b; Ingber, 1996b; Ingber, 1997; Ingber, 1998). This model was the first physical application of a nonlinear multivariate calculus developed by other mathematical physicists in the late 1970's (Graham, 1977; Langouche *et al*, 1982).

## 2.1. Statistical Aggregation

SMNI studies have detailed a physics of short-term memory and of (short-fiber contribution to) EEG phenomena (Ingber, 1984), in terms of $M^G$ firings, where $G$ represents $E$ or $I$, $M^E$ represents contributions to columnar firing from excitatory neurons, and $M^I$ represents contributions to columnar firing from inhibitory neurons. About 100 neurons comprise a minicolumn (twice that number in visual cortex); about 1000 minicolumns comprise a macrocolumn. A mesocolumn is developed by SMNI to reflect the convergence of short-ranged (as well as long-ranged) interactions of macrocolumnar input on minicolumnar structures, in terms of synaptic interactions taking place among neurons (about 10,000 synapses per neuron). The SMNI papers give more details on this derivation.

In this SMNI development, a Lagrangian is explicitly defined from a derived probability distribution of mesocolumnar firings in terms of the $M^G$ and electric potential variables, $\Phi^G$. Several examples have been given to illustrate how the SMNI approach is complementary to other models. For example, a mechanical string model was first discussed as a simple analog of neocortical dynamics to illustrate the general ideas of top down and bottom up interactions (Nunez, 1989; Nunez & Srinivasan, 1993). SMNI was applied is to this simple mechanical system, illustrating how macroscopic variables are derived from small-scale variables (Ingber & Nunez, 1990).

## 2.2. STM Stability and Duration

SMNI has presented a model of STM, to the extent it offers stochastic bounds for this phenomena during focused selective attention. This $7 \pm 2$ "rule" is well verified by SMNI for acoustical STM (Ingber, 1984; Ingber, 1985b; Ingber, 1994), transpiring on the order of tenths of a second to seconds, limited to the retention of $7 \pm 2$ items (Miller, 1956). The $4 \pm 2$ "rule" also is well verified by SMNI for visual or semantic STM, which typically require longer times for rehearsal in an hypothesized articulatory loop of individual items, with a capacity that appears to be limited to $4 \pm 2$ (Zhang & Simon, 1985). SMNI has detailed these constraints in models of auditory and visual cortex (Ingber, 1984; Ingber, 1985b; Ingber,

1994; Ingber & Nunez, 1995).

Another interesting phenomenon of STM capacity explained by SMNI is the primacy versus recency effect in STM serial processing (Ingber, 1985b), wherein first-learned items are recalled most error-free, with last-learned items still more error-free than those in the middle (Murdock, 1983). The basic assumption is that a pattern of neuronal firing that persists for many $\tau$ cycles, $\tau$ on the order of 10 msec, is a candidate to store the "memory" of activity that gave rise to this pattern. If several firing patterns can simultaneously exist, then there is the capability of storing several memories. The short-time probability distribution derived for the neocortex is the primary tool to seek such firing patterns. The highest peaks of this probability distribution are more likely accessed than the others. They are more readily accessed and sustain their patterns against fluctuations more accurately than the others. The more recent memories or newer patterns may be presumed to be those having synaptic parameters more recently tuned and/or more actively rehearsed.

It has been noted that experimental data on velocities of propagation of long-ranged fibers (Nunez, 1981; Nunez, 1995) and derived velocities of propagation of information across local minicolumnar interactions (Ingber, 1982) yield comparable times scales of interactions across minicolumns of tenths of a second. Therefore, such phenomena as STM likely are inextricably dependent on interactions at local and global scales.

### 2.3. SMNI Correlates of STM and EEG

Previous SMNI studies have detailed that maximal numbers of attractors lie within the physical firing space of $M^G$, consistent with experimentally observed capacities of auditory and visual short-term memory (STM), when a "centering" mechanism is enforced by shifting background noise in synaptic interactions, consistent with experimental observations under conditions of selective attention (Mountcastle *et al*, 1981; Ingber, 1984; Ingber, 1985b; Ingber, 1994; Ingber & Nunez, 1995). This leads to all attractors of the short-time distribution lying along a diagonal line in $M^G$ space, effectively defining a narrow parabolic trough containing these most likely firing states. This essentially collapses the 2 dimensional $M^G$ space down to a 1 dimensional space of most importance. Thus, the predominant physics of short-term memory and of (short-fiber contribution to) EEG phenomena takes place in a narrow "parabolic trough" in $M^G$ space, roughly along a diagonal line (Ingber, 1984).

Using the power of this formal structure, sets of EEG and evoked potential data, collected to investigate genetic predispositions to alcoholism, were fitted to an SMNI model to extract brain "signatures" of short-term memory (Ingber, 1997; Ingber, 1998). These results give strong quantitative support for an accurate intuitive picture, portraying neocortical interactions as having common algebraic or physics mechanisms that scale across quite disparate spatial scales and functional or behavioral phenomena, i.e., describing interactions among neurons, columns of neurons, and regional masses of neurons.

For future work, I have described how bottom-up neocortical models can be developed into eigenfunction expansions of probability distributions appropriate to describe short-term memory in the context of scalp EEG (Ingber, 2000). The mathematics of eigenfunctions are similar to the top-down eigenfunctions developed by some EEG analysts, albeit they have different physical manifestations. The bottom-up eigenfunctions are at the local mesocolumnar scale, whereas the top-down eigenfunctions are at the global regional scale. However, these approaches have regions of substantial overlap (Ingber & Nunez, 1990; Ingber, 1995a), and future studies may expand top-down eigenfunctions into the bottom-up eigenfunctions, yielding a model of scalp EEG that is ultimately expressed in terms of columnar states of neocortical processing of attention and short-term memory.

An optimistic outcome of future work might be that these EEG eigenfunctions, baselined to specific STM processes of individuals, could be a more direct correlate to estimates of *g* factors.

### 2.4. Complexity in EEG

In Chapter 6, Biological Correlates of *g*, AJ puts these issues in perspective:

First, psychometric tests were never intended or devised to measure anything other than behavioral variables. ... at this point most explanations are still conjectural.

However, most of the chapter falls back to similar too-simple statistical models of correlations between measured variables and behavioral characteristics.

The sections on Biological Correlates of *g* dealing with EEG recordings are incomplete and at a level too superficial relative most of the other parts of the book. The introduction of "complexity" as a possible correlate to IQ is based on faddish studies that do not have any theoretical or experimental support (Nunez, 1995). If such support does transpire, most likely it will be developed on the shoulders of more complete stochastic models and much better EEG recordings.

### 3. SMNI STM Correlates of the *g* Factor

### 3.1. High vs Low *g* Categories

The outline of test categories giving rise to high versus low *g* loadings on page 35 provides me with some immediate candidates for further investigations of a physiological basis for the *g* factor. I think a good working hypothesis is that these two categories are marked by having the high *g* loadings more correlated than the low *g* loadings to statistical interactions among the peaks of the probability distributions described above in the context of SMNI's theory of STM. The high *g* categories clearly require relatively more processing of several items of information than do the low *g* categories.

This seems to be a similar correlation as that drawn by Spearman, as described by AJ, in having "education of relations" and "education of correlates" more highly correlation with high *g* categories than the "apprehension of experience."

### 3.2. Mechanisms of High *g* Categories

From the SMNI perspective, control of selective attention generally is highly correlated with high utilization of STM, e.g., to tune the "centering mechanism." This seems similar to Spearman's correlation of "mental energy" with high *g* categories.

There are several mechanisms that might distinguish how well individuals might perform on high *g* category tasks. The ability to control the "centering mechanism" is required to sustain a high degree of statistical processes of multiple most probable states of information. The particular balance of general chemical-electrical activity directly shapes the distribution of most probable states. For some tasks, processing across relatively more most probable states might be required; for other tasks processing among larger focussed peaks of most probable states might be more important.

### 3.3. Hick's Law — Linearity of RT vs STM Information

The SMNI approach to STM gives a reasonable foundation to discuss RT and items in STM storage. These previous calculations support the intuitive description of items in STM storage as peaks in the 10-millisecond short-time (Ingber, 1984; Ingber, 1985b) as well as the several-second long-time (Ingber, 1994; Ingber & Nunez, 1995) conditional probability distribution of correlated firings of columns of neurons. These columnar firing states of STM tasks also were correlated to EEG observations of evoked potential activities (Ingber, 1997; Ingber, 1998). This distribution is explicitly calculated by respecting the nonlinear synaptic interactions among all possible combinatoric aggregates of columnar firing states (Ingber, 1982; Ingber, 1983).

The RT necessary to "visit" the states under control during the span of STM can be calculated as the mean time of "first passage" between multiple states of this distribution, in terms of the probability $P$ as an outer integral $\int dt$ (sum) over refraction times of synaptic interactions during STM time $t$, and an inner integral $\int dM$ (sum) taken over the mesocolumnar firing states $M$ (Risken, 1989), which has been explicitly calculated to be within observed STM time scales (Ingber, 1984),

$$RT = -\int dt\, t \int dM\, \frac{dP}{dt}\,. \tag{3}$$

As demonstrated by previous SMNI STM calculations, within tenths of a second, the conditional probability of visiting one state from another $P$, can be well approximated by a short-time probability distribution expressed in terms of the previously mentioned Lagrangian $L$ as

$$P = \frac{1}{\sqrt{(2\pi dtg)}} \exp(-Ldt)\,, \tag{4}$$

where $g$ is the determinant of the covariance matrix of the distribution $P$ in the space of columnar firings.

This expression for $RT$ can be approximately rewritten as

$$RT \approx K \int dt \int dM\, P \ln P\,, \tag{5}$$

where $K$ is a constant when the Lagrangian is approximately constant over the time scales observed. Since the peaks of the most likely $M$ states of $P$ are to a very good approximation well-separated Gaussian peaks (Ingber, 1984), these states by be treated as independent entities under the integral. This last expression is essentially the "information" content weighted by the time during which processing of information is observed.

The calculation of the heights of peaks corresponding to most likely states includes the combinatoric factors of their possible columnar manifestations as well as the dynamics of synaptic and columnar interactions. In the approximation that we only consider the combinatorics of items of STM as contributing to most likely states measured by $P$, i.e., that $P$ measures the frequency of occurrences of all possible combinations of these items, we obtain Hick's Law, the observed linear relationship of RT versus STM information storage, first discussed by AJ in Chapter 8. For example, when the bits of information are measured by the probability $P$ being the frequency of accessing a given number of items in STM, the bits of information in 2, 4 and 8 states are given as approximately multiples of $\ln 2$ of items, i.e., $\ln 2$, $2 \ln 2$ and $3 \ln 2$, resp. (The limit of taking the logarithm of all combinations of independent items yields a constant times the sum over $p_i \ln p_i$, where $p_i$ is the frequency of occurrence of item $i$.)

## 4. Conclusion

The book written by AJ provides motivation to explore a more fundamental basis of the *g* factor. I have examined this work in the narrow focus of some specific elements of SMNI, previously used to develop a theory of STM and a model of EEG.

I have focussed on how bottom-up SMNI models can be developed into eigenfunction expansions of probability distributions appropriate to describe STM. This permits RT to be calculated as an expectation value over the STM probability distribution of stored states, and in the good approximation of such states being represented by well separated Gaussian peaks, this yield the observed linear relationship of RT versus STM information storage.

This SMNI STM approach also suggests several other studies that can be performed in the context of examining an underlying basis for the *g* factor.

Basis for the *g* factor - 11 - Lester Ingber